# Improved FCM algorithm for Clustering on Web Usage Mining

K.Suresh[1] R.Madana Mohana[2] A.RamaMohanReddy[3]

[1] Department of Software Engineering, East China University of Technology,ECIT Nanchang Campus, Nanchang, Jiangxi-330013, P.R.China.

[2] Department of Information Technology, Vardhaman college of Engineering, Shamshabad, Hyderabad, A.P, India.

[3] Department of Computer Science and Engineering , S.V.University College of Engineering, Tirupati, A.P, India.

Abstract

In this paper we present clustering method is very sensitive to the initial center values ,requirements on the data set too high, and cannot handle noisy data the proposal method is using information entropy to initialize the cluster centers and introduce weighting parameters to adjust the location of cluster centers and noise problems. The navigation datasets which are sequential in nature, Clustering web data is finding the groups which share common interests and behavior by analyzing the data collected in the web servers, this improves clustering on web data efficiently using improved fuzzy c-means(FCM) clustering. Web usage mining is the application of data mining techniques to web log data repositories. It is used in finding the user access patterns from web access log. Web data Clusters are formed using on MSNBC web navigation dataset.

## 1. Introduction

The World Wide Web has huge amount information [1,2]and large datasets are available in databases. Information retrieving on websites is one of possible ways how to extract information from these datasets is to find homogeneous clusters of similar units for the description of the data vector descriptions are usually used each its component corresponds to a variable which can be measured in different scales (nominal, ordinal, or numeric) most of the well known clustering methods are implanted only for numeric data(k-means method) or are too complex for clustering large datasets(such as hierarchical methods based on dissimilarity matrices). Fuzzy clustering relevant for information retrieval as a document might be relevant to multiple queries, this document should be given in the corresponding response sets otherwise the user would not be aware of it ,Fuzzy clustering seems a natural technique for document categorization there are two basic methods of fuzzy clustering[4] ,one which is based on fuzzy c-partitions is called a fuzzy c-means clustering method and the other ,based on the fuzzy equivalence relations is called a fuzzy equivalence clustering method.

## 2. Clustering

Broadly speaking clustering algorithms[3] can be divided into two types partitioned and hierarchical. Partitioning algorithms construct a partition of a database D of n objects into a set of clusters where k is a input parameter.

Hierarchical algorithms create decomposition of the database D. they are a Agglomerative and divisive. Hierarchical clustering builds a tree of clusters, also known as a dendrogram. Every cluster node contains child cluster. An agglomerative clustering starts with one-point (singleton) Clusters and recursively merges two or more most appropriate clusters. A divisive clustering starts with one cluster of all data points and recursively splits into the most appropriate clusters. The process continues until a stopping criterion is achieved. There are two main issues in clustering techniques, Firstly finding the optimal number of clusters in a given dataset and secondly, given two sets of clusters, computing relative measure of goodness between them. For both these purposes, a criterion function or a validation function is usually applied. The simplest and most widely used cluster optimization function is the sum of squared error [5]. Studies on the sum of squared error clustering were focused on the well-known k-Means algorithm [6] and its variants.

In conventional clustering objects that are similar are allocated to the same cluster while objects that differ are put in different clusters. These clusters are hard clusters. In soft clustering an object may be in more than two or more clusters. Clustering is a widely used technique in data mining application for discovering patterns in underlying data. Most traditional clustering algorithms are limited in handling datasets that contain categorical attributes. However, datasets with categorical types of attributes are common in real life data mining problem. For each pair of documents, a comparison vector is constructed that contains binary features that measure the overlap for highly informative but sparse features between the two documents and numeric features.

The aggregating the comparison vector into one value that belongs to interval. The aggregation step is performed by taking a weighted average the information gain has a tendency to favor features with many possible values over feature with fewer possible values ,we used a normalized version of information gain, called gain ration as weighting metric.

Clustering is of prime importance in data analysis, machine learning and statistics. It is defines as the process of grouping N item sets into distinct clusters based on similarity or distance function .A good clustering technique may yield clusters thus have high inter cluster and low intra cluster distance[7].The objective of clustering is to maximize the similarity of the data points within each cluster and maximize dissimilarity across clusters.

Information gain of feature is calculated as follows. Assume we have K, the set of class label (a binary set in our case: document pairs belonging to the same cluster or not) and $M_i$, the set of feature values for feature I, with this information; we can calculate the database information entropy; the probabilities are estimated from the relative frequencies in the training set.

$$H(k) = -\sum_{k \in K} P(k)\log_2 P(k)$$

The information gain of feature i is then measured by calculating the difference in entropy between the situations with and without the information about the values of the feature.

$$W_i = H(k) - \sum_{m \in M_i} P(m) \times H(k/m)$$

gain ration is a normalized version of information gain. it is information gain divided by split info li(i),the entropy of the feature values. This I just the entropy of the database restricted to a single feature.

$$w_i = \frac{H(k) - \sum_{m \in M_i} P(m) \times H(K/m)}{li(i)}$$

$$li(i) = -\sum_{m \in M_i} P(m)\log_2 P(m)$$

## 3. Improved FCM clustering algorithm

As the FCM algorithm is very sensitive to the number of cluster centers, cluster centers initialization often artificially get significant errors, and even get the actual opposite results .FCM algorithm[8] is hard on data sets too ,so the data sets must be quite regular, in order to solve problems, first of all we use information entropy to initialize the cluster centers to determine the number of cluster centers. it can be reduce some errors, and also can improve the algorithm introductions an weighting parameters after that combine with the merger of ideas and divide the large chumps into small clusters. Then merge various small clusters according to the merger of the conditions, so that you can solve the irregular datasets clustering. Document similarity measures as shown in below.

The algorithm as follows

```
Initialize number of clusters
Initialize C_j (cluster centers)
Initialize α (threshold value)
Repeat
For i=1 to n :update μ_j(X_i)
        For k=1 to p ;
          Sum=0
          Count=0
        For i=1 to n:
        If μ (X_i) is maximum in C_k  then
         If μ (X_i)>= α
         Sum=sum+X_i
         Count= count+1
    C_k=sum/count
Until C_j estimate stabilize.
```

The clustering framing as follows
A set clusters $C=\{C_1, C_2, C_3, \ldots\ldots C_k\}$
Maximum precision values:

$$Purity= \sum_{i=1}^{k}\left(\frac{(|C_i|)}{n}\right) Max_{j=1}^{n} \Pr ecision(C_i, L_j)$$

$$Precision(C_i,L_j)= \left(\frac{(|C_i \cap L_j|)}{|C_i|}\right)$$

$$Inverse\ Purity= \sum_{i=1}^{m}\left(\frac{(|L_j|)}{n}\right) Max_{j=1}^{k} \operatorname{Re} call(C_i, L_j)$$

Recall(Cj,Li)=Precision(Li ,Cj)
To calculate the harmonic mean, the F-means

$$Purity\text{-}F= \sum_{i=1}^{m}\left(\frac{(|L_j|)}{n}\right) Max_{j=1}^{k} \{F(C_j, L_i)\}$$

Where the maximum is taken over all cluster F(Cj,Li) is defined as

$$F(Cj,Li)= \frac{2 \times \operatorname{Re} call(C_j, L_i) \times \Pr ecision(C_j, L_i)}{\operatorname{Re} call(C_{j,} L_i) + \Pr ecision(C_j, L_i)}$$

## Web usage data set

In this section we describe the dataset used and the description of the dataset used for the experimental results.

Information about the dataset

Number of users : 989818
Average number of visits per user : 5.7
Number of URLs per category : 0 to 5000

Table 1:Clustering of web usage data

| sequence | Order of user page visits |
|---|---|
| 1 | 1 1 |
| 2 | 2 |
| 3 | 3 2 2 4 2 2 2 3 3 |
| 4 | 5 |
| 5 | 1 |
| 6 | 6 |
| 7 | 1 1 |
| 8 | 6 |
| 9 | 6 7 7 7 6 6 8 8 8 8 |
| 10 | 6 9 4 4 4 10 3 10 5 10 4 4 4 |
| 11 | 1 1 1 11 1 1 1 |
| 12 | 12 12 |
| 13 | 1 1 |

Description of the Dataset we collected the data from the UCI dataset repository that consists of sever logs from msnbc.com for the month of September 1998. each sequence corresponds to page views of a user during that 24 hour period. Each sequence in the dataset corresponds to the page views of a user during that twenty four hour period. Each event in the sequence corresponds to a users request for a page.

There are 17 page categories "FrontPage", "news", "tech", "local", "opinion", "on-air", "misc", "weather", "health", "living", "business", "sports", "summary", "bbs" (bulletin board service), "travel", "msn-news", and "msn-sports".

Each category is associated in order with an integer starting with "1". For example, "FrontPage" is associated with 1, "news" with 2, and "tech" with 3. Each row below "% Sequences:" describes the hits in order of a single user. For example, the first user hits "FrontPage" twice, and the second user hits "news" once.

The length of the user sessions ranges from 1 to 500 and the average length of session is 5.7 .

Similarity matrix with p=0.5
Table 1:Clustering of web usage data

The first similarity upper approximation at threshold value 0.5 is given by

R(T1)={T1,T5,T7,T11,T13}
R(T2)={T2}
R(T3)={T3}
R(T4)={T4,}
R(T5)={T1,T5,T11,T13}
R(T6)={T6,T8}
R(T7)={T1,T7,T11,T13}
R(T8)={T6,T8}
R(T9)={T9}
R(T10)={T10}
R(T11)={T1,T5,T7,T11,T13}
R(T12)={T12}
R(T13)={T1,T5,T7,T11,T13}

Graphical representation of the clusters found after first upper approximations

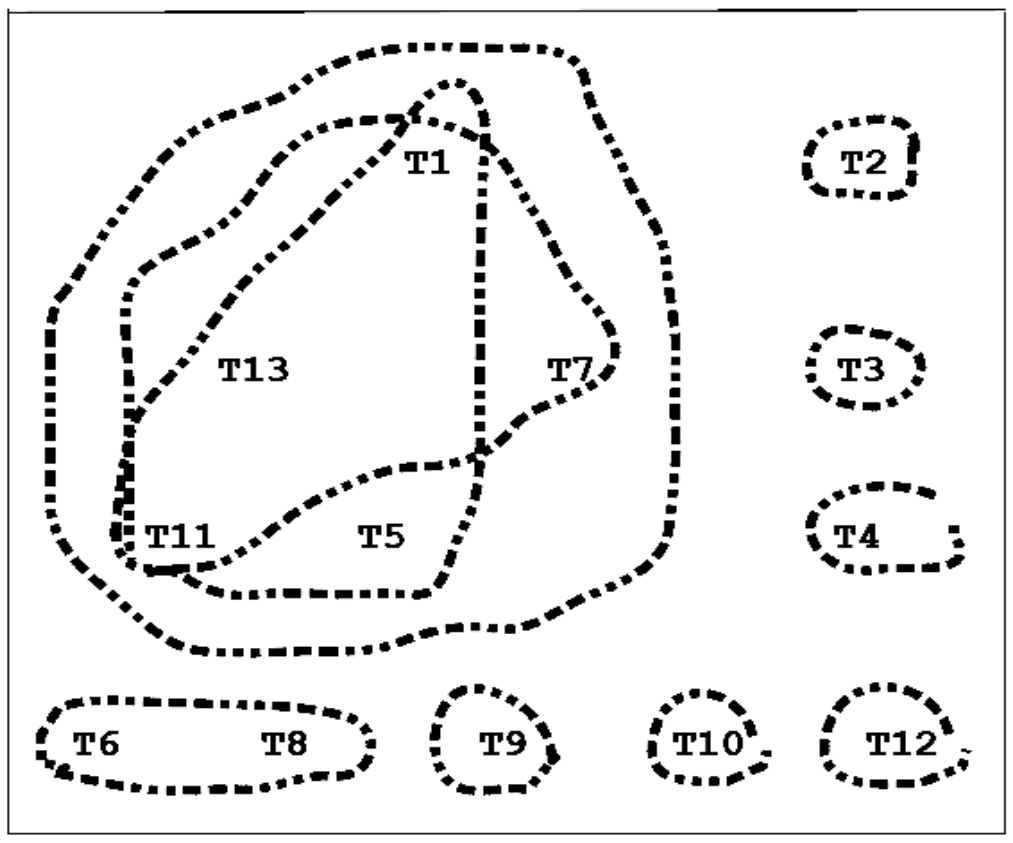

The similarity upper approximations for

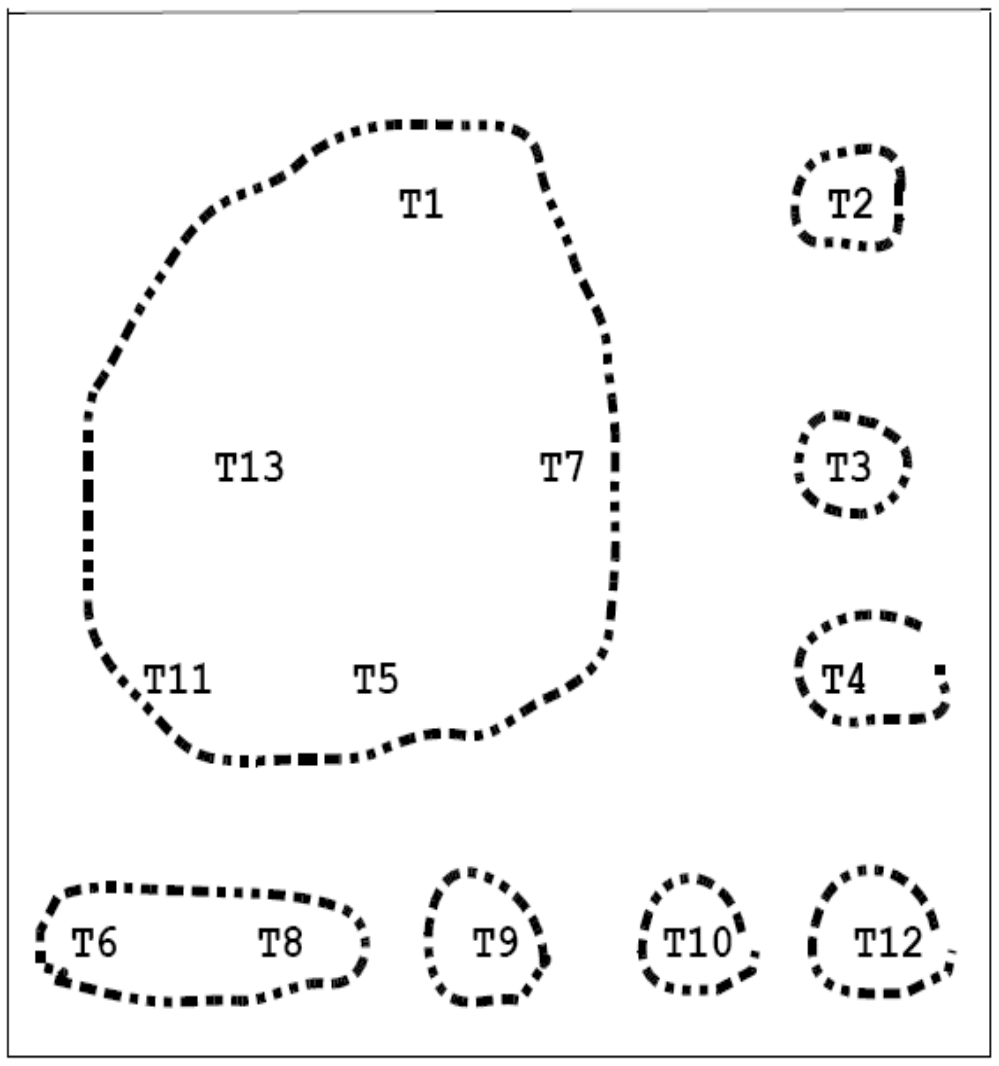

S1={ T1,T5,T7,T11, T13}
S2={T2,},S3={T3},S4={T4},S5={T1,T5,T7,T11,T13},S6 ={T6,T8},S7={T1,T5,T7,T11,T13},S8={T6,T8},S9={T9, },S10={T10},S11={T1,T5,T7,T11,T13},S12={T12},S13 ={T1,T5,T7,T11,T13}.
This denotes that user visiting the hyper links in t1 may visit the hyperlinks in t5 ad then t7, t11 and hyper links in t13.
One who visits the hyper links in t6 may also visit the hyper links in t8.

## 4. Conclusions

In this paper we presented a clustering web usage data ,from msnbc.com which is useful in finding the user access patterns and the order of visits of the hyperlinks of the each user. The suggested approach was used for efficacy contained a hard clustering of the msbn.com data set and as the analysis indicated each of the clusters seems to contain observations with specific common charterstics and improve the algorithm efficiency with help of improved FCM algorithm .Experiments prove the improved algorithm has able to identify the initial cluster centers.

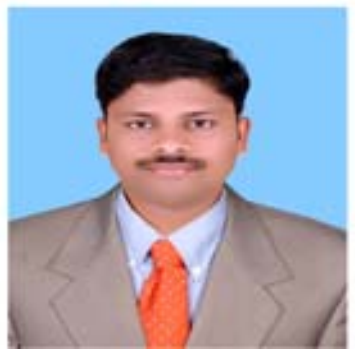

**K.Suresh** received his Bachelor degree and Master Degree in Information Technology from JNT University Hyderabad. He is currently working as Foreign Faculty in East China University of Technology, P.R.China and Visiting Faculty for Jiangxi Normal University, he worked in AITS, Rajampet, A.P, India .he has published more than 20 national and International conference papers and journals. He received best paper award in 2008 in National wide paper presentation. His field of interest is data mining, database technologies, information retrieval system.

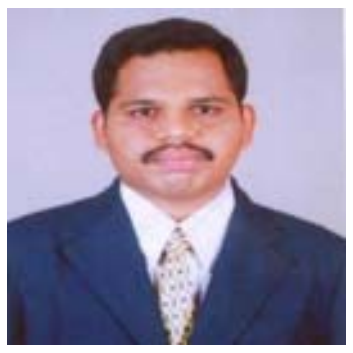

**R. Madana Mohana** isan Associate Professor in the Information Technology department at Vardhaman College of Engineering, Hyderabad, Andhra Pradesh, India since 2007. He has 8 years of teaching experience at both UG and PG levels. He received his B. Tech in Computer Science and Information Technology from Jawaharlal Nehru Technological University, Hyderabad in 2003 and M. E in Computer Science and Engineering from Sathyabama University ,Chennai in 2006. He is doing Ph. D in Computer Science and Engineering at Sri Venkateswara University Tirupathi, Andhra Pradesh, India. He is a life member of ISTE Technical Association. His areas of interest include Data Mining, Automata Theory, Compiler Design and Database Systems.

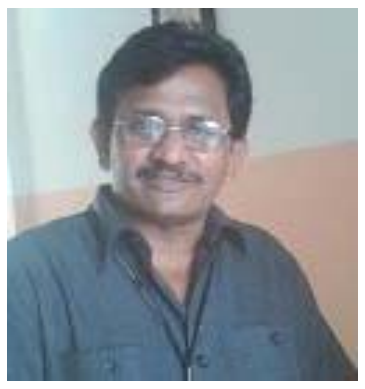

**Dr. A. Rama Mohan Reddy** received the B. Tech. from JNT University, Hyderabad in 1986, M. Tech degree in Computer Science from National Institute of Technology in 2000 Warangal and Ph. D in Computer Science and Engineering in 2008 from Sri Venkateswara University, Tirupathi, Andhra Pradesh, India. He worked as Assistant Professor, Associate Professor of Computer Science and Engineering, Sri Venkateswara University College of Engineering during the period 1992 and 2005. Presently working as Professor of Computer Science and Engineering, Sri Venkateswara University College of Engineering. He has 28 years of Industry and Teaching experience. Currently guiding twelve Ph. D scholars. He is life member of ISTE and IE. Research interests include Software Architecture, Software Engineering and Data Mining. He has 10 international publications and 14 international conference Publications at International and National level.